\documentclass[%
 reprint,
 superscriptaddress,
 amsmath,amssymb,
 aps,
 pra,
 longbibliography,
lengthcheck,%
]{revtex4-1}

\usepackage{graphicx}% Include figure files
\usepackage{dcolumn}% Align table columns on decimal point
\usepackage{bm}% bold math
\usepackage{hyperref}% add hypertext capabilities
\usepackage{color}
\usepackage{ulem}
\usepackage{xr}
\renewcommand{\Re}[0]{{\rm Re}}

\begin{document}

\preprint{APS/123-QED}

\title{
% Ferrimagnetic and antiferromagnetic insulators as a spin-current solar cell
Theory for shift current of bosons: Photogalvanic spin current\\ 
in ferrimagnetic and antiferromagnetic insulators
}
%optical は可視光ととられる

\author{Hiroaki Ishizuka}
\affiliation{
Department of Applied Physics, The University of Tokyo, Bunkyo, Tokyo, 113-8656, JAPAN 
}

\author{Masahiro Sato}
\affiliation{
Department of Physics, Ibaraki University,  Mito, Ibaraki, 310-8512, JAPAN
}

\date{\today}

\begin{abstract} 
We theoretically study the optical generation of dc spin current (i.e., a spin-current solar cell) in ordered antiferromagnetic and ferrimagnetic insulators, motivated by a recent study on the laser-driven spinon spin current in noncentrosymmetric quantum spin chains [H. Ishizuka and M. Sato, Phys. Rev. Lett. {\bf122}, 197702 (2019)]. Using a non-linear response theory for magnons, we analyze the dc spin current generated by a linearly-polarized electromagnetic wave (typically, terahertz or gigahertz waves). Considering noncentrosymmetric two-sublattice magnets as an example, we find a finite dc spin current conductivity at $T=0$, where no thermally-excited magnons exist; this is in contrast to the case of the spinon spin current, in which the optical transition of the Fermi degenerate spinons plays an essential role. We find that the dc spin-current conductivity is insensitive to the Gilbert damping, i.e., it may be viewed as a shift current carried by bosonic particles (magnons). Our estimate shows that an electric-field intensity of $E\sim10^4-10^6$ V/cm is sufficient for an observable spin current. Our theory indicates that the linearly-polarized electromagnetic wave generally produces a dc spin current in noncentrosymmetric magnetic insulators.
\end{abstract}

\maketitle

%%%%%%%%%%%%%%%%%%%%%%%%%%%%%%%%%%%%%%%%%%%%
%%%%%%%%%%%%%%%%%%%%%%%%%%%%%%%%%%%%%%%%%%%%
%%%%%%%%%%%%%%%%%%%%%%%%%%%%%%%%%%%%%%%%%%%%
\section{Introduction}

Materials subject to an intense incident light shows rich behaviors which are studied in the context of nonlinear response and non-equilibrium phenomena. An example of such is electric shift current in noncentrosymmetric semiconductors and ferroelectrics~\cite{Belinicher1982,Sturman1992,Sipe2000,Tan2016,Morimoto2016,Ogawa2017,Tokura2018}, where a non-trivial shift of electron position during its optical transition produces a macroscopic electric current. Recent studies revealed that the shift current exhibits strikingly different behaviors from the ordinary photocurrent; the shift current shows unique light-position dependence when it is excited locally~\cite{Ishizuka2017c,Nakamura2017,Bajpai2019}, and propagates faster than the Fermi velocity of electrons~\cite{Laman2005,Daranciang2012,Nakamura2017}. On the other hand, in correlated materials, lower-energy excitations often emerge due to the interaction effect; a typical example is magnetic excitations in Mott insulators. The optical transition of these emergent particles may produce non-trivial phenomena, especially, transport phenomena, related to the nonlinear response of the emergent excitations.

Several recent studies in opto-spintronics and magneto-optics~\cite{Kirilyuk2010,Nemec2018,Baltz2018} implies that the intensity and coherence of currently-available electromagnetic waves are sufficient for the control of magnetic excitations or magnetism. Typical results are the following: Magnetization switching by a circularly-polarized laser in ferrimagnets~\cite{Kimel2005,Stanciu2007,Khorsand2012,Hennecke2019}, 
laser-driven demagnetization~\cite{Beaurepaire1996,Koopmans2000,Stamm2007}, the spin pumping by gigahertz (GHz) or terahertz (THz) waves~\cite{Kajiwara2010,Heinrich2011}, focused-laser driven magnon propagation~\cite{Satoh2012,Hashimoto2017}, 
intense THz-laser driven magnetic resonance~\cite{Mukai2016,Lu2017}, spin control by THz-laser driven electron transitions~\cite{Baierl2018}, dichroisms driven by THz vortex beams~\cite{Cheong2019}, 
angular momentum transfer between photons and magnons in cavities~\cite{Haigh2015,Osada2016,Zhang2016,Haigh2016,Usami2018}, a ultrafast detection of spin Seebeck effect~\cite{Kampfrath2018}, a phonon-mediated spin dynamics with THz laser~\cite{Kampfrath2018b}, etc. Moreover, recent theoretical works have proposed several ways of optical control of magnetism: THz-wave driven inverse Faraday effect~\cite{Takayoshi2014-1,Takayoshi2014-2}, Floquet engineering of magnetic states such as chirality ordered states~\cite{Sato2016,Kitamura2017} and a spin liquid state~\cite{Sato2014}, generation of magnetic defects with laser-driven heat~\cite{Koshibae2014,Fujita2017-1}, applications of topological light waves to magnetism~\cite{Fujita2017-1,Fujita2017-2,Fujita2018-1,Fujita2018-2}, control of exchange couplings in Mott insulators with high-~\cite{Mentink2015} and low-frequency~\cite{Takasan2018} waves, optical control of spin chirality in multiferroic materials~\cite{Mochizuki2010}, rectification of dc spin currents in magnetic insulators with electromagnetic waves~\cite{Proskurin2018,Ishizuka2019,Okuma2019}. These studies are partly supported by recent developments in THz laser science~\cite{JPhysD2017,Cavalleri2017} which realized high-intensity light beams with the photon energy comparable to those of magnetic excitations. Despite these developments, the optical control of the current carried by magnetic excitations is limited to some theoretical proposals. 

Among the proposals, a recent theory proposes a mechanism for producing a dc spin current in quantum spin chains without the angular-momentum transfer~\cite{Ishizuka2019}; it is distinct from the known mechanisms in which the angular momentum of photons are transferred to the magnet~\cite{Kajiwara2010,Heinrich2011,Ohnuma2014,Proskurin2018,Okuma2019}. The mechanism in Ref.~\onlinecite{Ishizuka2019} is analogous to that of the shift-current photovoltaic effect~\cite{Sturman1992}. The close relation between two phenomena are clear from the Jordan-Wigner fermion representation of spin chain; the ground state of the spin chain is a band insulator of Jordan-Wigner fermions, and the photovoltaic response is related to the optical transition of the fermions by the linearly-polarized THz light. However, the relation of this mechanism to the fermion excitations casts doubt on the generality because the low-energy excitations of the ordered magnets are usually magnons, i.e., bosonic excitations.

In this work, we theoretically show that a dc spin current similar to that of the spin chain~\cite{Ishizuka2019} also appears in ordered antiferromagnetic (AFM) and ferrimagnetic (FRM) insulators by applying a linearly-polarized electromagnetic wave. The symmetry argument in Sec.~\ref{sec:Zeeman1d} shows that the creation of dc spin current with linearly-polarized waves is possible only if both site- and bond-center inversion symmetries are broken. AFM and FRM insulators violate the bond-center inversion symmetry and thereby they naturally satisfy half of the required symmetry condition. The staggered moment is an advantage of considering AFM/FRM insulators for generating a dc spin current. As an example, we consider two-sublattice models with N\'eel type ground state. Bosonic particles describe the low-energy excitations of these models, i.e., magnons; the ground state is the zero-magnon state. This ground state is very different from that of noncentrosymmetric $S=1/2$ spin chains~\cite{Ishizuka2019} which are described by a Fermi degenerated state of spinons. Despite the difference, our calculation using a nonlinear response theory finds a finite photovoltaic spin current similar to that of the spinons. We discuss that it is related to the zero-point fluctuation of the quantum magnets. Our theory also indicates that the magnon spin current is shift-current like, i.e., it is insensitive to the magnon lifetime as in the spinon case. This mechanism allows generation of spin current using a linearly-polarized electromagnetic wave and ordinary AFM or FRM insulators.

The remaining part of the paper is organized as follows. In Sec.~\ref{sec:KvB}, we introduce the nonlinear-response theory for two-species magnons, which we will use in the following sections. The main results of this paper are in Secs.~\ref{sec:Zeeman1d} and \ref{sec:Zeeman3d}. 
Section~\ref{sec:Zeeman1d} focuses on the photo-induced spin current in AFM and FRM insulators 
with a strong one dimensionality, while we study the three-dimensional (3D) magnets in Sec.~\ref{sec:Zeeman3d}.   
Effective experimental setups and signatures for investigating the proposed mechanism are discussed in Sec.~\ref{sec:experiment}. Section~\ref{sec:summary} is devoted to the summary and discussions.

%%%%%%%%%%%%%%%%%%%%%%%%%%%%%%%%%%%%%%%%%%%%
%%%%%%%%%%%%%%%%%%%%%%%%%%%%%%%%%%%%%%%%%%%%
%%%%%%%%%%%%%%%%%%%%%%%%%%%%%%%%%%%%%%%%%%%%
\section{Nonlinear response theory}\label{sec:KvB}

We calculate the nonlinear response coefficients for the photo-induced spin current by extending the linear-response theory to the quadratic order in the perturbation. A similar method for fermions is used to calculate the photovoltaic current in semiconductors~\cite{Kraut1979,vBaltz1981} and the spin current of spinons~\cite{Ishizuka2019}. The derivation of the formula is summarized in Appendix~\ref{sec:ResponseEq}. We here summarize the outline of the derivation. We also discuss the physical implications.

We consider a two-sublattice AFM/FRM insulator with two species of magnons. The effective Hamiltonian for the magnons is 
\begin{align}
H=\sum_{\bm k} \varepsilon_\alpha(\bm k)\alpha^\dagger_{\bm k}\alpha_{\bm k}+\varepsilon_\beta(\bm k)\beta_{\bm k}\beta^\dagger_{\bm k},\label{eq:KvB:H0}
\end{align}
where $\alpha_{\bm k}$ ($\alpha_{\bm k}^\dagger$) and $\beta_{\bm k}$ ($\beta_{\bm k}^\dagger$) are the boson annihilators (creators) for the magnons with the momentum $\bm k=(k_x,k_y,k_z)$ and $\varepsilon_{a}(\bm k)$ ($a=\alpha,\beta$) is the energy of the magnons in the $a=\alpha,\beta$ branch with momentum $\bm k$. We here consider a general perturbation (spin-electromagnetic-wave coupling)
\begin{align}
H'=&-\sum_{\mu,\bm k}\int \frac{d\omega}{2\pi}\,h^\mu_\omega e^{{\rm i}\omega t}
\psi_{\bm k}^\dagger
\left(\begin{array}{cc}
(B_{\bm k}^\mu)_{\alpha\alpha} & (B_{\bm k}^\mu)_{\alpha\beta} \\
(B_{\bm k}^\mu)_{\beta\alpha} & (B_{\bm k}^\mu)_{\beta\beta} 
\end{array}\right)
\psi_{\bm k}\nonumber\\
&\qquad\qquad\qquad+\text{h.c.},
\end{align}
and spin-current operator
\begin{align}
J=&\sum_{\bm k}
\psi_{\bm k}^\dagger
\left(\begin{array}{cc}
(J_{\bm k})_{\alpha\alpha} & (J_{\bm k})_{\alpha\beta} \\
(J_{\bm k})_{\beta\alpha} & (J_{\bm k})_{\beta\beta} 
\end{array}\right)
\psi_{\bm k}.
\end{align}
Here, $\omega$ is the frequency of ac light, $h^\mu_\omega$ is the spin-light coupling constant for the $\mu$ direction, 
and $\psi_{\bm k}=(\alpha_{\bm k},\beta_{-\bm k}^\dagger)^T$.

The nonlinear conductivity is defined by
\begin{align}
\langle J\rangle(\Omega)=\sum_{\mu,\nu}\int d\omega\,\sigma_{\mu\nu}(\Omega;\omega,\Omega-\omega)h^{\mu}_\omega h^{\nu}_{\Omega-\omega},
\end{align}
where $\langle J\rangle(\Omega)\equiv\int dt\langle J\rangle(t)e^{-{\rm i}\Omega t}$ is the Fourier transform of the expectation value of the spin current $\langle J\rangle(t)$.
For the two-sublattice model, the formula for nonlinear spin current conductivity reads
\begin{widetext}
\begin{align}
\sigma_{\mu\nu}(\Omega;\omega,\Omega-\omega)=&\frac1{2\pi} \sum_{\bm k,a_i=\alpha,\beta}\frac{{\rm sgn}(a_3)(\tilde\rho_{\bm k,a_1}{\rm sgn}(a_2)-{\rm sgn}(a_1)\tilde\rho_{\bm k,a_2})(B^\mu_{\bm k})_{a_1a_2}}{\omega-\tilde\varepsilon_{a_2}(\bm k)+\tilde\varepsilon_{a_1}(\bm k)-i/(2\tau_{\bm k})}\nonumber\\
&\hspace{2cm}\times\left[\frac{(B^\nu_{\bm k})_{a_2a_3}(J_{\bm k})_{a_3a_1}}{\Omega+\tilde\varepsilon_{a_1}(\bm k)-\tilde\varepsilon_{a_3}(\bm k)-i/(2\tau_{\bm k})}-\frac{(J_{\bm k})_{a_2a_3}(B^\nu_{\bm k})_{a_3a_1}}{\Omega+\tilde\varepsilon_{a_3}(\bm k)-\tilde\varepsilon_{a_2}(\bm k)-i/(2\tau_{\bm k})}\right],
\label{eq:sigmagen}
\end{align}
\end{widetext}
where
\begin{align}
\tilde\varepsilon_{a}(\bm k)=&{\rm sgn}(a)\varepsilon_{a}(\bm k),\\
{\rm sgn}(a)=&\left\{\begin{array}{rl}
1 & (a=\alpha)\\
-1& (a=\beta)
\end{array}\right.,\\
\tilde\rho_{\bm k,a}=&\left\{\begin{array}{rl}
\langle\alpha^\dagger_{\bm k}\alpha_{\bm k}\rangle_0 & (a=\alpha)\\
\langle\beta_{-\bm k}\beta_{-\bm k}^\dagger\rangle_0 & (a=\beta)
\end{array}\right..
\end{align}
The relaxation time of magnons, $\tau_{\bm k}$, was introduced in Eq.~(\ref{eq:sigmagen}), and $\langle\cdots\rangle_0$ is the expectation value of $\cdots$ in the equilibrium state of the Hamiltonian in Eq.~\eqref{eq:KvB:H0}.
The conductivity for dc spin current corresponds to the $\Omega=0$ case, $\sigma_{\mu\nu}(0;\omega,-\omega)$. In the rest of this work, we focus on the case $B_{\bm k}^\mu=B_{\bm k}^\nu=B_{\bm k}$ because we are interested in the response to a linearly polarized light. Hence, we abbreviate the subscripts in the nonlinear conductivity, $\sigma_{\mu\nu}(0;\omega,-\omega)=\sigma(0;\omega,-\omega)$.
%%%%%%%%%%%%%%%%%%%%%%%%%%%%%%%%%%%%%

\begin{figure}
  \includegraphics[width=\linewidth]{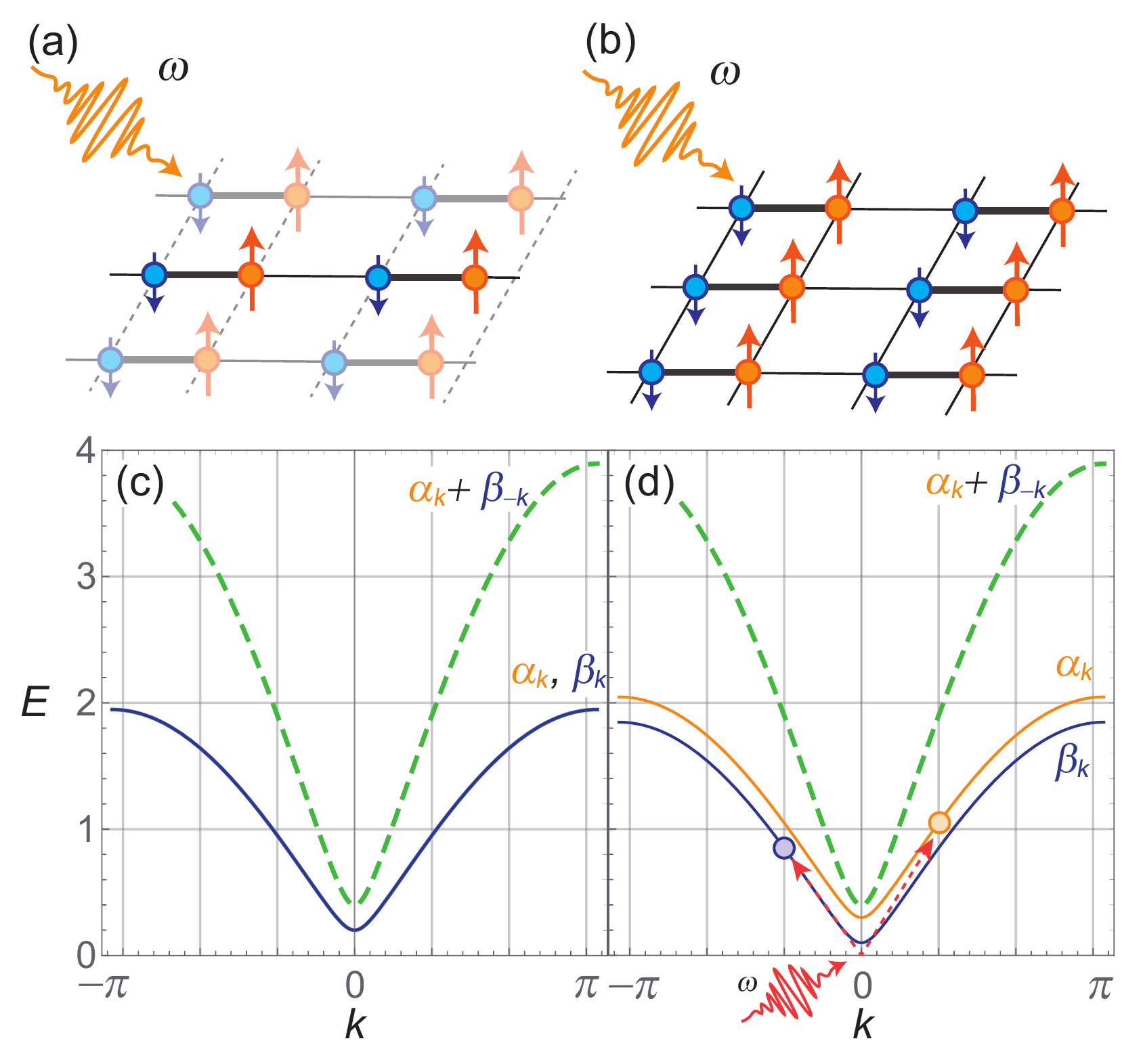}
  \caption{(Color online) Schematic pictures of the noncentrosymmetric magnets. A quasi-one-dimensional magnet consisting of weakly-coupled spin chains (a) and a three-dimensional magnet with two-sublattice order (b). Each sublattice (blue and orange) has a different environment, e.g., different $g$ factors, uniaxial anisotropy, etc., and with the bond dimerization (shown by the thick bond). The two-sublattice order and bond dimerization respectively breaks the inversion symmetry on the bond center and sites. Magnetic excitation and nonlinear spin current conductivity of the spin chain. The magnon band dispersions of the model in Eq.~\eqref{eq:H1} for (c) $h_+=0$ and (d) $h_+=1/100$. Parameters $h_\pm$ are defined in Eq.~(\ref{eq:h_pm}). When $h_+=0$, two magnon dispersions are degenerate. The THz light produces two magnons, one on each branch as schematically shown in panel (d).}
  \label{fig:model}
\end{figure}

%%%%%%%%%%%%%%%%%%%%%%%%%%%%%%%%%%%%%%%%%%

We note that the conductivity in Eq.~\eqref{eq:sigmagen} remains non-zero at $T=0$. The substitutions of $\tilde\rho_{\bm k,\alpha}=0$ and $\tilde\rho_{\bm k,\beta}=1$ in Eq.~\eqref{eq:sigmagen} reduce the formula to
\begin{align}
\sigma&(0;\omega,-\omega)=\sum_{\bm k}\nonumber\\
-&\frac1\pi\left[\frac{(1+i2\tau_{\bm k}\omega)|(B_{\bm k})_{\beta\alpha}|^2((A_{\bm k})_{\alpha\alpha}+(A_{\bm k})_{\beta\beta})}{(\omega-i/2\tau_{\bm k})^2-(\varepsilon_{\alpha}(\bm k)+\varepsilon_{\beta}(\bm k))^2}\right]\nonumber\\
+&\frac1{2\pi}\frac{(B_{\bm k})_{\beta\alpha}(A_{\bm k})_{\alpha\beta}((B_{\bm k})_{\beta\beta}+(B_{\bm k})_{\alpha\alpha})}{(\omega-i/2\tau_{\bm k}-\varepsilon_\alpha(\bm k)-\varepsilon_\beta(\bm k))(\varepsilon_\alpha(\bm k)+\varepsilon_\beta(\bm k)+i/2\tau_{\bm k})}\nonumber\\
+&\frac1{2\pi}\frac{(B_{\bm k})_{\alpha\beta}(A_{\bm k})_{\beta\alpha}((B_{\bm k})_{\beta\beta}+(B_{\bm k})_{\alpha\alpha})}{(\omega-i/2\tau_{\bm k}+\varepsilon_\alpha(\bm k)+\varepsilon_\beta(\bm k))(\varepsilon_\alpha(\bm k)+\varepsilon_\beta(\bm k)-i/2\tau_{\bm k})}.\label{eq:sigmagen2}
\end{align}
Because of $\tilde\rho_{\bm k,\beta}=1$, the terms involving the off-diagonal component of $B_{\bm k}$ remains at $T=0$. In other words, the two-magnon creation/annihilation process plays a crucial role as shown in Fig.~\ref{fig:model}(d). We focus on the $T=0$ case in the rest of this paper as this process is dominant in the low temperature limit.

From a different viewpoint, Eq.~\eqref{eq:sigmagen2} implies the zero-point fluctuation plays a key role in the photovoltaic response of magnons. In our formalism, the zero-point fluctuation is manifested in the Bogoliubov transformation of Holstein-Primakov bosons. This transformation creates $\beta_{\bm k}\beta^\dagger_{\bm k}$ and $\alpha^\dagger_{\bm k}\beta^\dagger_{-\bm k}$ terms which contribute to the photovoltaic response in the ground state. $\tilde\rho_{\bm k,\beta}=1$ is another consequence of the Bogoliubov transformation. The importance of the zero-point fluctuation resembles the spinon spin current~\cite{Ishizuka2019}, in which the Fermi degeneracy of spinons represents the quantum fluctuation of spins. A crucial difference in the current case is the absence of Fermi degeneracy. However, in the case of the AFMs/FRMs, the condensate of Holstein-Primakov bosons plays a similar role to the Fermi degeneracy. The pair-creation process represented by $\alpha_{\bf k}^\dagger \beta_{\bf k}^\dagger$ generates photovoltaic response of the magnons which is manifested in the denominator of Eq.~\eqref{eq:sigmagen2}; the sum of eigenenergies, $\varepsilon_\alpha(\bm k)+\varepsilon_\beta(\bm k)$, represents creation/annihilation of a magnon pair. These features implies that the zero-point fluctuation is necessary for the shift current response at $T=0$.

The first term in Eq.~\eqref{eq:sigmagen2} vanishes when the ground state has a certain symmetry. For example, collinear magnetic orders with the moments parallel to $S^z$ axis are often symmetric with respect to $G={\cal T}M_x^s$, which is the product of time-reversal operation (${\cal T}$) and the mirror operation for the spin degrees of freedom about $x$ axis ($M_x^s$). In this case, the real part of $\sigma(0;\omega,-\omega)$ reads
\begin{align}
&\Re\left[\sigma(0;\omega,-\omega)\right]=\nonumber\\
&-\frac1{\pi}\sum_{\bm k}\Re\left\{\frac{(B_{\bm k})_{\beta\alpha}(A_{\bm k})_{\alpha\beta}((B_{\bm k})_{\beta\beta}+(B_{\bm k})_{\alpha\alpha})}{\omega^2-(\varepsilon_\alpha(\bm k)+\varepsilon_\beta(\bm k)+i/2\tau_{\bm k})^2}\right\}.\label{eq:sigmagen3}
\end{align}
The conductivities for the models considered in the following sections are calculated using this formula.

%%%%%%%%%%%%%%%%%%%%%%%%%%%%%%%%%%%%%%%%%%%%
%%%%%%%%%%%%%%%%%%%%%%%%%%%%%%%%%%%%%%%%%%%%
%%%%%%%%%%%%%%%%%%%%%%%%%%%%%%%%%%%%%%%%%%%%
\section{Spatially-anisotropic magnet}\label{sec:Zeeman1d}

In this section, we apply the above formula to a spin chain with AFM or FRM order, which corresponds to a quasi-one-dimensional (quasi-1D) magnetic compound with a negligible inter-chain interaction. The spins are coupled to the electromagnetic wave through the Zeeman coupling. To make the problem theoretically well-defined, we consider a model which conserves the spin angular momentum $S^z$; the model has an easy axis and the applied ac magnetic field is parallel to the ordered moments. The conservation of $S^z$ allows us to unambiguously define the spin current operator from the continuity equation. This setup is in contrast to those of usual magnetic resonances and spin pumping~\cite{Kajiwara2010,Heinrich2011}, in which the ac field is perpendicular to the magnetic moment.  
We use the standard spin-wave approximation to describe magnetic excitations (magnons).

%%%%%%%%%%%%%%%%%%%%%%%%%%%%%%%%%%%%%%%%%%%%
%%%%%%%%%%%%%%%%%%%%%%%%%%%%%%%%%%%%%%%%%%%%
%%%%%%%%%%%%%%%%%%%%%%%%%%%%%%%%%%%%%%%%%%%%
\subsection{Model}
We consider an ordered noncentrosymmetric spin chain with a two-sublattice unit cell [Fig.~\ref{fig:model}(a)], whose Hamiltonian is given by 
\begin{subequations}
\begin{align}
H_{\rm tot}=&H_0+H_Z^{(\omega)}\label{eq:H1},\\
H_0\equiv&\sum_{r_y,r_z}H_{\rm 1D}(r_y,r_z),\\
H_{\rm 1D}(r_y,r_z)\equiv&\sum_{r_x}J(1+\delta)\bm S_A(\bm r)\cdot\bm S_B(\bm r)\nonumber\\
&+J(1-\delta)\bm S_A(\bm r+\hat x)\cdot\bm S_B(\bm r)\nonumber\\
&-(D+D_s) \left[S_A^z(\bm r)\right]^2-(D-D_s)\left[S_B^z(\bm r)\right]^2\nonumber\\
&\qquad-h\left[g_AS_A^z(\bm r)+g_BS_B^z(\bm r)\right],\label{eq:H1dZ}\\
H_Z^{(\omega)}=&-(h_\omega e^{{\rm i}\omega t}+\text{h.c.})\sum_{\bm r} g_AS_A^z(\bm r)+g_BS_B^z(\bm r),
\end{align}
\end{subequations}
where $H_{\rm 1D}$ is the spin-chain Hamiltonian with the staggered nearest-neighbor exchange interaction (i.e., dimerization) along the $x$ direction, $H_0$ is the bundle of all the chains, and $H_Z^{(\omega)}$ is the Zeeman coupling between the spins and the external electromagnetic wave. Here, $\bm S_a(\bm r)\equiv(S_a^x(\bm r),S_a^y(\bm r),S_a^z(\bm r))$ ($a=A,B$) is the spin-$S_a$ operator on the $a$ sublattice of the unit cell at position $\bm r=(r_x,r_y,r_z)$. 
Symbols $\hat x$, $\hat y$, and $\hat z$ stand for the unit vectors along the $x$, $y$, and $z$ directions, respectively. 
The parameters in the Hamiltonian $H_{\rm 1D}$ are as follows: $J>0$ is the antiferromagnetic exchange interaction 
along the spin-chain ($x$) direction, $\delta$ is the dimerization, $D>0$ ($D_s$) is the uniform easy-axis (staggered) anisotropy, $g_A$ ($g_B$) is the $g$ factor for the spins on $A$ ($B$) sublattice, and $h$ is the external static magnetic field along the $S^z$ axis. 
In the spin-light coupling $H_Z^{(\omega)}$, $|h_\omega|$ and $\arg(h_\omega)$ %in Eq.~\eqref{eq:H1dZ} 
are respectively the magnitude and the phase of the ac magnetic field of the linearly-polarized electromagnetic wave. 
We assume %the size of spins on $a=A,B$ sublattice is $S_a$, 
$|D_s|<D$, and $|\delta|<1$.

When $S_A\ne S_B$, the ground state of the model in Eq.~\eqref{eq:H1dZ} is a FRM-ordered state with magnetization $|S_A-S_B|$ per a unit cell~\cite{Marshall1955,Lieb1962}. The ground state is a N\'eel ordered when $S_A=S_B$. The classical ground state of $H_0$ has a collinear order with spins pointing along the $S^z$ axis because of the easy axis anisotropy $D$ [Fig.~\ref{fig:model}(a)]. The anisotropy also produces the spin gap in the excitation spectrum [Fig.~\ref{fig:1dchain}(a) and 1(b)]. We discuss the effect of the gap and its relation to the frequency dependence of the nonlinear spin conductivity in the next section. 

Here we define the spin current for $S^z$. Since the model $H_0$ conserves the $z$ component of total spin angular momentum, the spin current for $S^z$ can be defined from the continuity equations $\partial_t S_A^z=J_x^z(r_x-1,B;r_x,A)-J_z^z(r_x,A;r_x,B)$ and $\partial_t S_B^z=J_x^z(r_x,A;r_x,B)-J_z^z(r_x,B;r_x+1,A)$, in which $J_\beta^\alpha(r,a;r',b)$ is the local spin-$S^\alpha$ current operator between two neighboring sites $(r,a)$ and $(r',b)$ and it flows along the $\beta$ direction. The above continuity equation is obtained from Heisenberg equation of motion for local spins. With these procedures, we find the uniform current operator for $H_{\rm 1D}$ reads
\begin{align}
J^z_x=& \frac{J}{2N}\sum_{\bm r}(1+\delta)\left\{S^x_B(\bm r)S^y_A(\bm r)-S^y_B(\bm r)S^x_A(\bm r)\right\}\nonumber\\
&+(1-\delta)\left\{S^x_A(\bm r+\hat x)S^y_B(\bm r)-S^y_A(\bm r+\hat x)S^x_B(\bm r)\right\},
\end{align}
where $N$ is the total number of unit cells.

%%%%%%%%%%%%%%%%%%%%%%%%%%%%%%%%%%%%%%%%%%%%
%%%%%%%%%%%%%%%%%%%%%%%%%%%%%%%%%%%%%%%%%%%%
%%%%%%%%%%%%%%%%%%%%%%%%%%%%%%%%%%%%%%%%%%%%
\subsection{Linear spin-wave approximation}\label{sec:spinwave}

Hereafter, we assume that in the ground state of $H_{\rm 1D}$, the spins on the $A$ sublattice points up 
while those on $B$ sublattice are down [see Fig.~\ref{fig:model} (a)].
The low-energy excitations of $H_0$ is calculated by linear spin-wave approximation. 
Using the Holstein-Primakov bosons, the spin operators are given by
\begin{subequations}
\begin{align}
S_A^z=&S_A-\hat n_A(\bm r),\\
S_A^+(\bm r)=&\sqrt{2S_A}\left(1-\frac{\hat n_A(\bm r)}{2S_A}\right)^{\frac12}a(\bm r),\\
S_A^-=&\sqrt{2S_A}a^\dagger(\bm r)\left(1-\frac{\hat n_A(\bm r)}{2S_A}\right)^{\frac12},
\end{align}
\end{subequations}
for the $A$ sublattice and
\begin{subequations}
\begin{align}
S_B^z=&\hat n_B(\bm r)-S_B,\\
S_B^+(\bm r)=&\sqrt{2S_B}b^\dagger(\bm r)\left(1-\frac{\hat n_B(\bm r)}{2S_B}\right)^{\frac12},\\
S_B^-=&\sqrt{2S_B}\left(1-\frac{\hat n_B(\bm r)}{2S_B}\right)^{\frac12}b(\bm r),
\end{align}
\end{subequations}
for the $B$ sublattice. Up to the linear order in $S_A$ and $S_B$, $H_0$ reads
\begin{align}
H_0\sim&\sum_{\bm k}\left(\begin{array}{c}
a_{\bm k} \\
b_{-\bm k}^\dagger
\end{array}\right)^\dagger
\left(\begin{array}{cc}
h^0_k+h^z_k & h^x_k-{\rm i}h^y_k \\
h^x_k+{\rm i}h^y_k & h^0_k-h^z_k
\end{array}\right)
\left(\begin{array}{c}
a_{\bm k} \\
b_{-\bm k}^\dagger
\end{array}\right)\nonumber\\
&+\text{const.}
\label{eq:MagH}
\end{align}
where the wave number along the chain ($x$) direction is simply represented by $k$, $a_{\bm k}\equiv(1/\sqrt{N})\sum_{\bm r} a(\bm r)e^{{\rm i}\bm k\cdot\bm r}$, $b_{\bm k}\equiv(1/\sqrt{N})\sum_{\bm r} b(\bm r)e^{{\rm i}\bm k\cdot(\bm r+\hat x/2)}$ are the Fourier transformation of Holstein-Primakov bosons. 
The matrix elements of the magnon Hamiltonian (\ref{eq:MagH}) are calculated as  
\begin{subequations}
\begin{align}
h^0_k=&h_++J(S_A+S_B),\\
h^x_k=&2J\sqrt{S_AS_B}\cos(k/2),\\
h^y_k=&-2J\delta\sqrt{S_AS_B}\sin(k/2),\\
h^z_k=&h_--J(S_A-S_B),
\end{align}
\label{eq:elements}
\end{subequations}
where
\begin{subequations}
\begin{align}
h_+=&D(S_A+ S_B-1)+D_s(S_A-S_B)+\frac{h}2(g_A-g_B),\\
h_-=&D(S_A-S_B)+D_s(S_A+S_B-1)+\frac{h}2(g_A+g_B).
\end{align}
\label{eq:h_pm}
\end{subequations}

\begin{figure}
  \includegraphics[width=\linewidth]{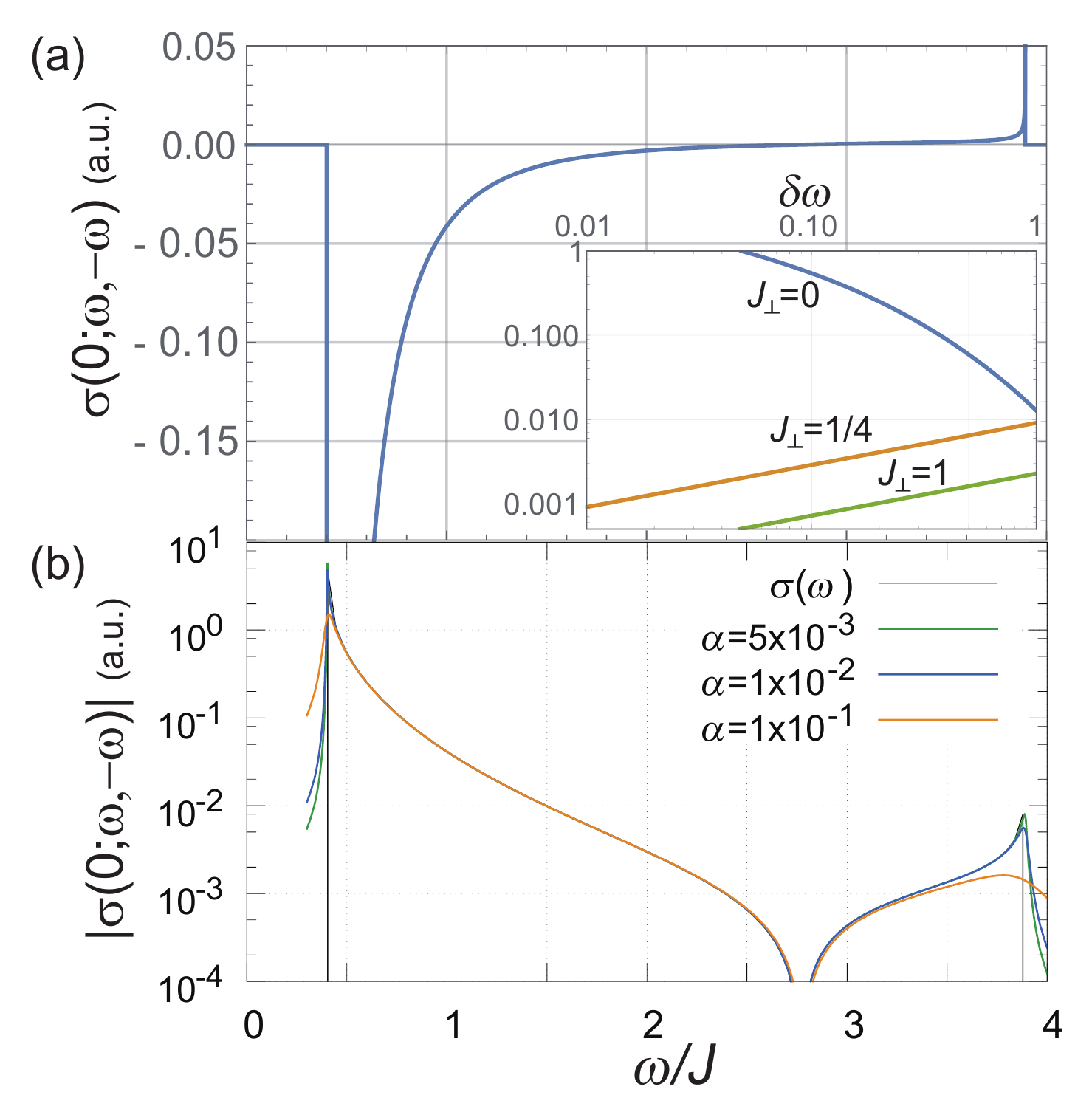}
  \caption{(Color online) Frequency dependence of the non-linear spin current conductivity$\sigma(0;\omega,-\omega)$. (a) Analytic result for the small Gilbert damping limit $\alpha\to0$ and (b) numerical results for a finite $\alpha$. The inset in (a) is the $\delta\omega\equiv\omega-\omega_{c1}$ for different $J_\perp$. The calculations are done using a chain with $N=2048-32768$ unit cells. All results are for $J=1$, $\delta=1/4$, $S_A=S_B=1$, $g_A=1$, $g_B=1/2$, $h_-=0$, and $h_+=1/100$ unless noted explicitly.}
  \label{fig:1dchain}
\end{figure}

The quadratic Hamiltonian (\ref{eq:MagH}) is diagonalized by the Bogoliubov transformation:
\begin{align}
a_{\bm k}=&\cosh\Theta_{\bm k}\alpha_{\bm k}+\sinh\Theta_{\bm k}\beta^\dagger_{-\bm k},\\
b_{-\bm k}^\dagger=&\sinh\Theta_{\bm k}e^{i\Phi_{\bm k}}\alpha_{\bm k}
+\cosh\Theta_{\bm k}e^{i\Phi_{\bm k}}\beta_{-\bm k}^\dagger, 
\label{eq:model:Bogoliubov}
\end{align}
where $\alpha_{\bm k}$ ($\alpha_{\bm k}^\dagger$) and $\beta_{\bm k}$ ($\beta_{\bm k}^\dagger$) are bosonic annihilation (creation) operators. By choosing
\begin{subequations}
\begin{align}
e^{{\rm i}\Phi_{\bm k}}=\frac{h^x_k+{\rm i}h^y_k}{\sqrt{(h^x_{k})^2+(h^y_{k})^2}}
\end{align}
and
\begin{align}
\cosh(2\Theta_{\bm k})=&\frac{h^0_{k}}{\sqrt{(h^0_{k})^2-(h^x_{k})^2-(h^y_{k})^2}},\\
\sinh(2\Theta_{\bm k})=&-\frac{\sqrt{(h^x_{k})^2+(h^y_{k})^2}}{\sqrt{(h^0_{k})^2-(h^x_{k})^2-(h^y_{k})^2}},
\label{eq:model:Theta}
\end{align}
\end{subequations}
the Hamiltonian becomes
\begin{align}
H_0=&\sum_{\bm k}\varepsilon_\alpha(\bm k)\alpha_{\bm k}^\dagger\alpha_{k}
+\varepsilon_\beta(\bm k)\beta_{-\bm k}^\dagger\beta_{-\bm k},
\end{align}
where
\begin{subequations}
\begin{align}
\varepsilon_\alpha(\bm k)=&h^z_{k}+\sqrt{(h^0_{k})^2-(h^x_{k})^2-(h^y_{k})^2},\\
\varepsilon_\beta(\bm k)=&-h^z_{k}+\sqrt{(h^0_{k})^2-(h^x_{k})^2-(h^y_{k})^2}.
\end{align}
\end{subequations}
Here, we ignored the constant term in $H_0$. We note that the dispersions $\varepsilon_{\alpha,\beta}(\bm k)$ and 
the phases $(\Theta_{\bm k},\Phi_{\bm k})$ are all independent of $k_y$ and $k_z$ because we now consider 
the 1D model $H_0$. Using the same transformation, we find
\begin{align}
H_Z^{(\omega)}
=&h\sum_{\bm k} (g_A\cosh^2\Theta_{\bm k}-g_B\sinh^2\Theta_{\bm k})\alpha_{\bm k}^\dagger\alpha_{\bm k}\nonumber\\
&+(g_A\sinh^2\Theta_{\bm k}-g_B\cosh^2\Theta_{\bm k})\beta_{-\bm k}\beta_{-\bm k}^\dagger\nonumber\\
&+\frac{g_A-g_B}2\sinh(2\Theta_{\bm k})(\alpha_{\bm k}^\dagger\beta_{-\bm k}^\dagger+\beta_{-\bm k}\alpha_{\bm k})\nonumber\\
&+g_B+hN(g_BS_B-g_AS_A).
\end{align}
and
\begin{widetext}
\begin{align}
J^z_x=&J\sqrt{S_AS_B}\sum_{\bm k}\sinh(2\Theta_{\bm k})\left(\sin\frac{k}2\cos\Phi_{\bm k}+\delta\cos\frac{k}2\sin\Phi_{\bm k}\right)\left(\alpha_{\bm k}^\dagger\alpha_{\bm k}+\beta_{-\bm k}\beta_{-\bm k}^\dagger\right)\nonumber\\
&\qquad+\left[\left\{\cosh(2\Theta_{\bm k})\left(\cos\Phi_{\bm k}\sin\frac{k}2-\delta\sin\Phi_{\bm k}\cos\frac{k}2\right)+{\rm i}\left(\sin\Phi_{\bm k}\sin\frac{k}2+\delta\cos\Phi_{\bm k}\cos\frac{k}2\right)\right\}\alpha^\dagger_{\bm k}\beta^\dagger_{-\bm k}+\text{h.c.}\right].\label{eq:model:Jdef}
\end{align}

%%%%%%%%%%%%%%%%%%%%%%%%%%%%%%%%%%%%%%%%%%%%
%%%%%%%%%%%%%%%%%%%%%%%%%%%%%%%%%%%%%%%%%%%%
%%%%%%%%%%%%%%%%%%%%%%%%%%%%%%%%%%%%%%%%%%%%
\subsection{Spin current conductivity}
Combining the magnon representation of $(\alpha_{\bm k},\beta_{\bm k})$ with the formula~(\ref{eq:sigmagen3}), 
we compute the nonlinear dc spin-current conductivity for the model $H_{\rm tot}$ under the application of THz laser. We first study the nonlinear conductivity in the clean limit with infinite relaxation time $\tau_k\to\infty$. The analytic solution for the conductivity $\Re\left[\sigma(0;\omega,-\omega)\right]$ obtained from Eq.~\eqref{eq:sigmagen3} reads
\begin{align}
\Re\left[\sigma(0;\omega,-\omega)\right]
=&\frac{(g_A-g_B)^2\delta(h_++J(S_A+S_B))\left(\omega^2-4(h_++J(S_A+S_B))^2-2J^2S_AS_B(1+\delta^2)\right)}{8\pi(1-\delta^2)\omega^2\sqrt{4J^4S_A^2S_B^2(1-\delta^2)^2-\{(\omega/4)^2+2J^2S_AS_B(1+\delta^2)-(h_++J(S_AS_B))^2\}^2}},\label{eq:sigma1d}
\end{align}
\end{widetext}
when $\omega\in[\omega_{c1},\omega_{c2}]$ and zero otherwise. Here,
\begin{align}
\omega_{c1}\equiv&\varepsilon_{\alpha}(0)+\varepsilon_{\beta}(0)\nonumber\\
=&2\sqrt{(h_++J(S_A+S_B))^2-4J^2S_AS_B},
\end{align}
corresponds to the energy for the band bottom of the pair excitation and
\begin{align}
\omega_{c2}\equiv&\varepsilon_{\alpha}(\pi)+\varepsilon_{\beta}(\pi)\nonumber\\
=&2\sqrt{(h_++J(S_A+S_B))^2-4\delta^2J^2S_AS_B},
\end{align}
is that for the top of the pair excitation [See Fig.~\ref{fig:model}(c) and (d)]. The frequency dependence of the conductivity is shown in Fig.~\ref{fig:1dchain}(a).
%%%%%%%%%%%%%%%%%%%%%%%%%%%%%%%%%%%%%

%%%%%%%%%%%%%%%%%%%%%%%%%%%%%%%%%%%%%%%%%%
Equation~\eqref{eq:sigma1d} is an odd function of $\delta$. This reflects the fact that the inversion-symmetry breaking is necessary for the spin current. $H_0$ has two inversion centers when $\delta=0$, $D_s=0$, $g_A=g_B$, and $S_A=S_B$: one at the center of the bond and the another on the site. The inversion center on the site is broken by the dimerization $\delta$. To see the dependence of $\sigma(0;\omega,-\omega)$ on the model parameters, we explicitly write down the nonlinear conductivity as a function of the parameters, i.e., $\sigma(0;\omega,-\omega)=\sigma(\omega;\delta,g_A-g_B,D_s,m)$, where $m=\langle S^z_{\bm r\in A}\rangle-\langle S^z_{\bm r\in B}\rangle$ is the order parameter of the AFM or FRM insulators. A symmetry argument on the transport coefficient finds $\sigma(\omega;\delta,g_A-g_B,D_s,m)=-\sigma(\omega;-\delta,g_A-g_B,D_s,m)$ for the site-center inversion operation. This result is identical to the spinon case in Ref.~\cite{Ishizuka2019}.

On the other hand, the magnetic order changes the parameter dependence of $\sigma(0;\omega,-\omega)$, which is related to the bond-center inversion operation. The inversion operation about the center of the bonds is broken by the N\'eel ordering, $D_s\ne0$ or $g_A\ne g_B$. Therefore, the symmetry operation indicates $\sigma(\omega;\delta,g_A-g_B,D_s,m)=-\sigma(\omega;\delta,-g_A+g_B,-D_s,-m)$. In addition, the translation operation about half a unit cell switches $A$ and $B$ sublattices and $m\to-m$; $\sigma(\omega;\delta,g_A-g_B,D_s,m)=-\sigma(\omega;\delta,g_A-g_B,D_s,m)$. Hence, the conductivity in the ordered phase is an even function of $g_A-g_B$ and $D_s$. This is a different behavior from the spinon case, in which the conductivity is an odd function of the staggered magnetic field (corresponds to $g_A-g_B$ in our case).

The conductivity diverges when $\omega$ approaches $\omega_{c1}$. The asymptotic form reads
\begin{align}
\Re&\left[\sigma(0;\omega,-\omega)\right]\approx\nonumber\\
&-\frac{(g_A-g_B)^2J^2\delta S_AS_B(h_++J(S_A+S_B))}{8\pi J\{(h_++J(S_A+S_B))^2-4J^2S_AS_B\}^{\frac54}}\nonumber\\
&\times\frac{1}{\sqrt{(1-\delta^2)S_AS_B\delta\omega}},
\end{align}
where $\delta\omega\equiv\omega-\omega_{c1}$. A similar feature is also found in the spinon case, in which the divergence is related to the singularity of the density of states~\cite{Ishizuka2019}.
On the other hand, the asymptotic form around $\omega=\omega_{c2}$ reads
\begin{align}
\Re&\left[\sigma(0;\omega,-\omega)\right]\approx\nonumber\\
&\frac{(g_A-g_B)^2J^2\delta S_AS_B (h_++J(S_A+S_B))}{8\pi J\{(h_++J(S_A+S_B))^2-4J^2\delta^2S_AS_B\}^{\frac54}}\nonumber\\
&\quad\times\frac{1}{\sqrt{(1-\delta^2)S_AS_B|\delta\omega|}}.
\end{align}
The sign of the conductivity is the opposite of that in the lower frequency regime. This is in contrast to the spinon case~\cite{Ishizuka2019}, in which the sign of the nonlinear conductivity remains the same for all frequencies $\omega\in[\omega_{c1},\omega_{c2}]$.

%\begin{figure}
%  \includegraphics[width=\linewidth]{figC.pdf}
%  \caption{(Color online) Result of numerical calculation with different Gilbert damping $\alpha$. We consider a chain with $S_A=S_B=1$, $g_a=1$, $g_B=1/2$, $\delta=1/4$, $h_+=1/100$, and $T=10^{-6}$. The calculations are done using a chain with $N=2048-32768$ unit cells. The thin black line is the result for $\alpha=0$.}\label{fig:gilbert}
%\end{figure}

%%%%%%%%%%%%%%%%%%%%%%%%%%%%%%%%%%%%%%%%%%%%
%%%%%%%%%%%%%%%%%%%%%%%%%%%%%%%%%%%%%%%%%%%%
%%%%%%%%%%%%%%%%%%%%%%%%%%%%%%%%%%%%%%%%%%%%
\subsection{Relaxation-time dependence}

We next study the damping (relaxation time) dependence of the spin current. In the study of photovoltaic effect, the relaxation-time dependence reflects the microscopic mechanism behind the photovoltaic effect~\cite{Sturman1992}; it is called shift current when $\sigma(0;\omega,-\omega)\propto \tau^0$ while is an injection current when $\sigma(0;\omega,-\omega)\propto\tau$. In bosonic systems, a slight difference appears in the momentum dependence of the single-particle relaxation time~\cite{Abrikosov1965}; it is inversely proportional to the momentum for the Goldstone modes. Therefore, we assume the momentum dependence of damping term as $\tau_{\bm k}=1/(\alpha_0 \varepsilon_\beta(\bm k))$ so that the momentum dependence is consistent with the field theoretic requirement ($\alpha_0$ is the damping factor). 
Physically, the assumed form of $\tau_{\bm k}$ corresponds to the phenomenological Gilbert damping. 
%We used Eq.~\eqref{eq:formula:BdG} for the calculation.

We substitute $\tau_{\bm k}=1/(\alpha_0 \varepsilon_\beta(\bm k))$ in Eq.~(\ref{eq:sigmagen3}) in order to estimate the relaxation-time dependence of the spin-current conductivity. 
Figure~\ref{fig:1dchain}(b) shows the $\alpha_0$ dependence of $\sigma(0;\omega,-\omega)$. Our numerical result shows $\sigma(0;\omega,-\omega)$ is insensitive to the damping. A slight difference, however, appears in the high-frequency region, where the smearing due to the damping is more distinct than that in the low-frequency region. This behavior is related to the momentum dependence of $\tau_k$, which is inversely proportional to the energy of the magnon. The insensitivity is a signature of a shift-current type photo-induced current~\cite{Sturman1992}; this is a similar feature to the spinon case~\cite{Ishizuka2019}.

%%%%%%%%%%%%%%%%%%%%%%%%%%%%%%%%%%%%%%%%%%%%
%%%%%%%%%%%%%%%%%%%%%%%%%%%%%%%%%%%%%%%%%%%%
%%%%%%%%%%%%%%%%%%%%%%%%%%%%%%%%%%%%%%%%%%%%
\section{Three-dimensional magnetic insulators}\label{sec:Zeeman3d}

\begin{figure}
  \includegraphics[width=0.8\linewidth]{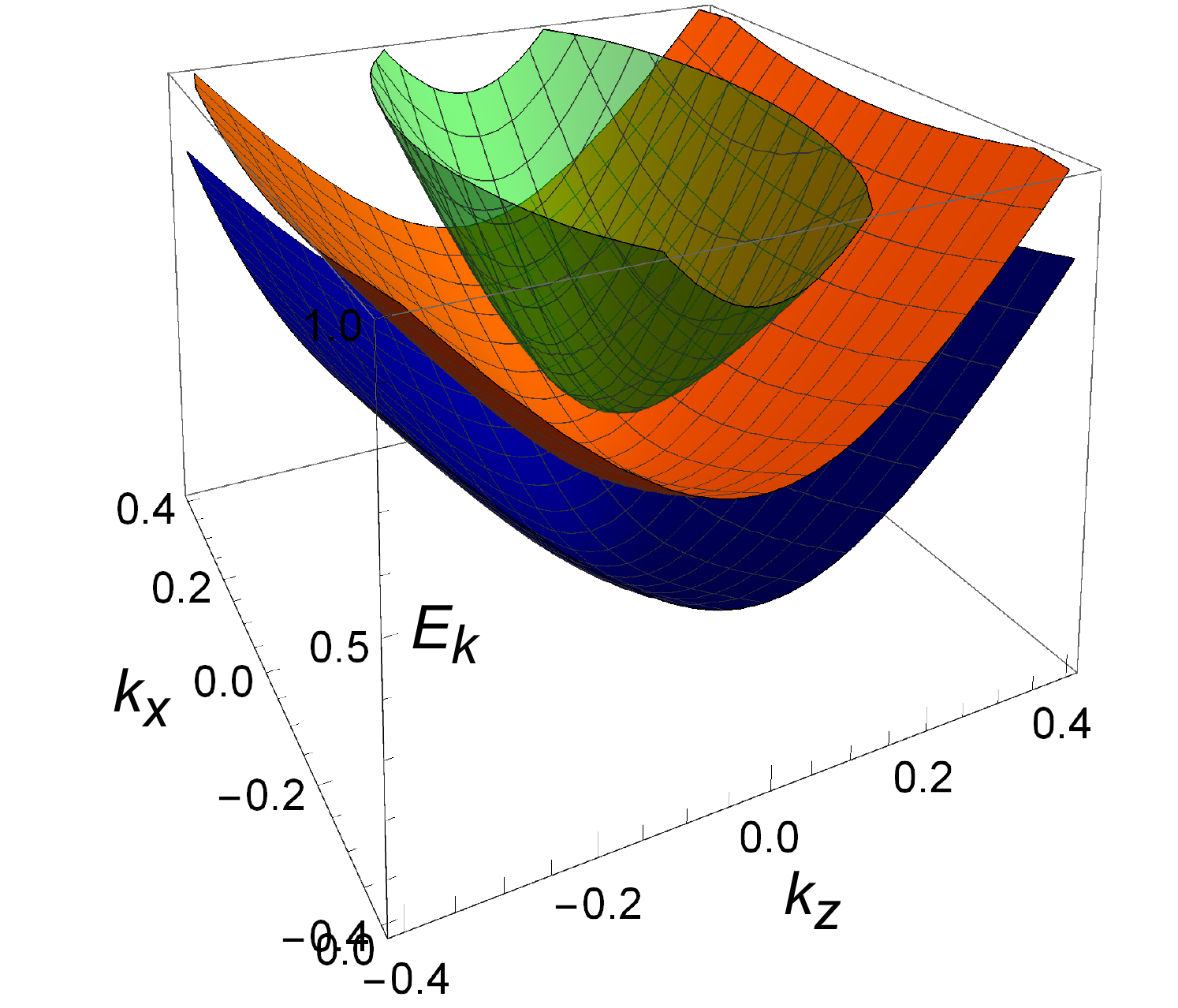}
  \caption{(Color online) Schematic figure of the magnon dispersion for the 3D model in Eq.~\eqref{eq:model3d}. We set $k_y=0$. The blue and orange planes are the dispersions of two magnon branches and the green transparent plane is that of two-magnon excitation. The plot is for $J=1$, $J_\perp=1$, $\delta=1/4$, $S_A=1$, $S_B=1$, $h_+=1/100$, and $h_-=1/10$.}\label{fig:model3d}
\end{figure}

In this section, we consider a three-dimensional (3D) magnet which consists of coupled spin chains $H_{\rm 1D}$ 
with a non-negligible inter-chain interaction [See Fig.~\ref{fig:model}(b)]. We particularly focus on the limit in which $\omega$ is close to the band gap of two-magnon excitations. The procedure of the calculation is the same as the 1D case in the previous section.
The static part of the Hamiltonian reads
\begin{align}
H_0^{(3D)}\equiv&\sum_{\bm r} J(1+\delta)\bm S_A(\bm r)\cdot\bm S_B(\bm r)\nonumber\\
&+J(1-\delta)\bm S_A(\bm r+\hat x)\cdot\bm S_B(\bm r)\nonumber\\
&-(D+D_s) \left[S_A^z(\bm r)\right]^2-(D-D_s)\left[S_B^z(\bm r)\right]^2\nonumber\\
&-J_\perp\left[\bm S_A(\bm r)\cdot\bm S_A(\bm r+\hat y)+\bm S_A(\bm r)\cdot\bm S_A(\bm r+\hat z)\right.\nonumber\\
&\qquad\left.+\bm S_B(\bm r)\cdot\bm S_B(\bm r+\hat y)+\bm S_B(\bm r)\cdot \bm S_B(\bm r+\hat z)\right]\nonumber\\
&-h\left[g_AS_A^z(\bm r)+g_BS_B^z(\bm r)\right].\label{eq:model3d}
\end{align}
The spin chains is parallel to the $x$ direction, while the $y$ and $z$ directions are perpendicular to the chains. 
The ferromagnetic coupling $J_\perp>0$ denotes the strength of the inter-chain exchange interaction. 
We study this model within the linear spin-wave approximation using Holsten-Primakov transformation in Sec.~\ref{sec:spinwave}. Focusing on the lower edge of the magnon dispersion, 
we first expand the matrix elements $h_{\bm k}^a$ of the magnon Hamiltonian [See Eq.~(\ref{eq:elements})] 
up to second order in $\bm k$:
\begin{subequations}
\begin{align}
h_{\bm k}^0\simeq&h_++J(S_A+S_B)+\frac{J_\perp(S_A+S_B)}2(k_y^2+k_z^2),\\
h_{\bm k}^x\simeq&J\sqrt{S_AS_B}(2-\frac14k_x^2),\\
h_{\bm k}^y\simeq&-J\sqrt{S_AS_B}\delta k_x,\\
h_{\bm k}^z\simeq&h_-+J(S_A-S_B)+\frac{J_\perp(S_A-S_B)}2(k_y^2+k_z^2).
\end{align}
\end{subequations}
We note that the magnon dispersions depend on both intra- and inter-chain wave numbers differently from the 1D case. 
The dispersion around $\Gamma$ point $\bm k=\bm 0$ is shown in Fig.~\ref{fig:model3d}. 
Using the momentum gradient of the low-energy Hamiltonian with $h_{\bm k}^{0,x,y,z}$, we can define the spin current operator; this approximation is essentially equivalent to expanding the lattice spin-current operator in Eq.~\eqref{eq:model:Jdef} up to the linear order in $\bm k$:
\begin{align}
J_z^z=&J\sqrt{S_AS_B}\sum_{\bm k}\sinh(2\Theta_{\bm k})\left(\frac{k_x}2\cos\Phi_{\bm k}+\delta\sin\Phi_{\bm k}\right)\nonumber\\
&\qquad\qquad\qquad\times\left(\alpha_{\bm k}^\dagger\alpha_{\bm k}+\beta_{-\bm k}\beta_{-\bm k}^\dagger\right)\nonumber\\
&\qquad+\left\{\cosh(2\Theta_{\bm k})\left(\cos\Phi_{\bm k}\frac{k_x}2-\delta\sin\Phi_{\bm k}\right)\right.\nonumber\\
&\qquad\qquad\left.+{\rm i}\left(\frac{k_x}2\sin\Phi_{\bm k}+\delta\cos\Phi_{\bm k}\right)\right\}\alpha^\dagger_{\bm k}\beta^\dagger_{-\bm k}+\text{h.c.}.
\end{align}
These equations corresponds to the $k\cdot p$ expansion of the lattice model. Therefore, it should be a good approximation for the lattice model when $\omega$ is close to the gap for two-magnon excitations.

The spin current conductivity is calculated using the formula of Eq.~\eqref{eq:sigmagen3}. 
A calculation similar to the 1D model considered in Sec.~\ref{sec:Zeeman1d} gives
\begin{align}
\Re&\left[\sigma(0;\omega,-\omega)\right]=\nonumber\\
&-\frac{J^2\delta S_AS_B(g_A-g_B)^2}{(4\pi)^22J_\perp\omega^2(S_A+S_B)}\left(8k_x-k_x^3\right)_{k_x=K_X},
\end{align}
where
\begin{align}
K_X=&%\left[8(1-\delta^2)\right.\nonumber\\
\Bigg[8(1-\delta^2)\nonumber\\
  -4&\left.\sqrt{\frac{(h_++J(S_A+S_B))^2-(\omega/2)^2}{J^2S_AS_B}+\delta^2(\delta^2-2)}\right]^{\frac12}.
\end{align}
When $\omega$ is close to the lower edge, i.e.,
\begin{align}
\omega\sim \omega_{c1}\equiv 2\sqrt{(h_++J(S_A+S_B))^2-4J^2S_AS_B},
\end{align}
$K_X$ becomes
\begin{align}
K_X\approx \sqrt{\frac{2\sqrt{(h_++J(S_A+S_B))^2-4J^2S_AS_B}\delta\omega}{(1-\delta^2)J^2S_AS_B}},
\end{align}
where $\delta\omega=\omega-\omega_{c1}$. Therefore, the asymptotic form of $\Re\left[\sigma_{ABB}(0;\omega,-\omega)\right]$ is
\begin{align}
\Re&\left[\sigma(0;\omega,-\omega)\right]\approx-\frac{(g_A-g_B)^2\delta\sqrt{S_AS_B}}{16\pi^2\sqrt{1-\delta^2}(S_A+S_B)}\nonumber\\
&\times\frac{J\sqrt{\delta\omega}}{J_\perp\left\{(h_++J(S_A+S_B))^2-4J^2S_AS_B\right\}^{\frac34}}.\label{eq:sigma3d}
\end{align}

Unlike the 1D case, in which the conductivity diverges at the band edge $\omega_{c1}$, the 3D result in Eq.~\eqref{eq:sigma3d} decreases proportionally to $\sqrt{\delta\omega}$ when approaching $\omega_{c1}$. The result is plotted in the inset of Fig.~\ref{fig:1dchain}(a) with the results for the 1D limit. This difference is a consequence of the difference in the density of states: it diverges in the 1D model while it is proportional to $\sqrt{\delta\omega}$ in the present 3D case.

The approximation we used in this section is accurate when $\omega$ is close to the magnon gap at the $\Gamma$ point in the Brillouin zone. In our model, the band bottom for the two-magnon excitations are at $\Gamma$ point, and the bandwidth of two-magnon excitation along the $x$ and $y$ directions are in the order of $J_\perp$ and that for $z$ direction is in the order of $J$. Therefore, our approximation is accurate when $\delta\omega \ll J, J_\perp$. This condition is manifested in $J_\perp$ in the denominator of Eq.~\eqref{eq:sigma3d}, which implies the divergence of $\Re\left[\sigma(0;\omega,-\omega)\right]$ at $J_\perp\to0$. When $J_\perp$ is very small, we expect $\Re\left[\sigma(0;\omega,-\omega)\right]$ to behave like that of the 1D case. On the other hand, $\Re\left[\sigma(0;\omega,-\omega)\right]$ looks like Eq.~\eqref{eq:sigma3d} when $J_\perp$ is sufficiently large, e.g., when $J_\perp\sim J$. Therefore, the 1d result and the result in this section corresponds to the two limits of the 3D magnet.

%%%%%%%%%%%%%%%%%%%%%%%%%%%%%%%%%%%%%%%%%%%%
%%%%%%%%%%%%%%%%%%%%%%%%%%%%%%%%%%%%%%%%%%%%
%%%%%%%%%%%%%%%%%%%%%%%%%%%%%%%%%%%%%%%%%%%%
\section{Experimental observation}\label{sec:experiment}

%%%%%%%%%%%%%%%%%%%%%%%%%%%%%%%%%%%%%

\begin{figure}
  \includegraphics[width=\linewidth]{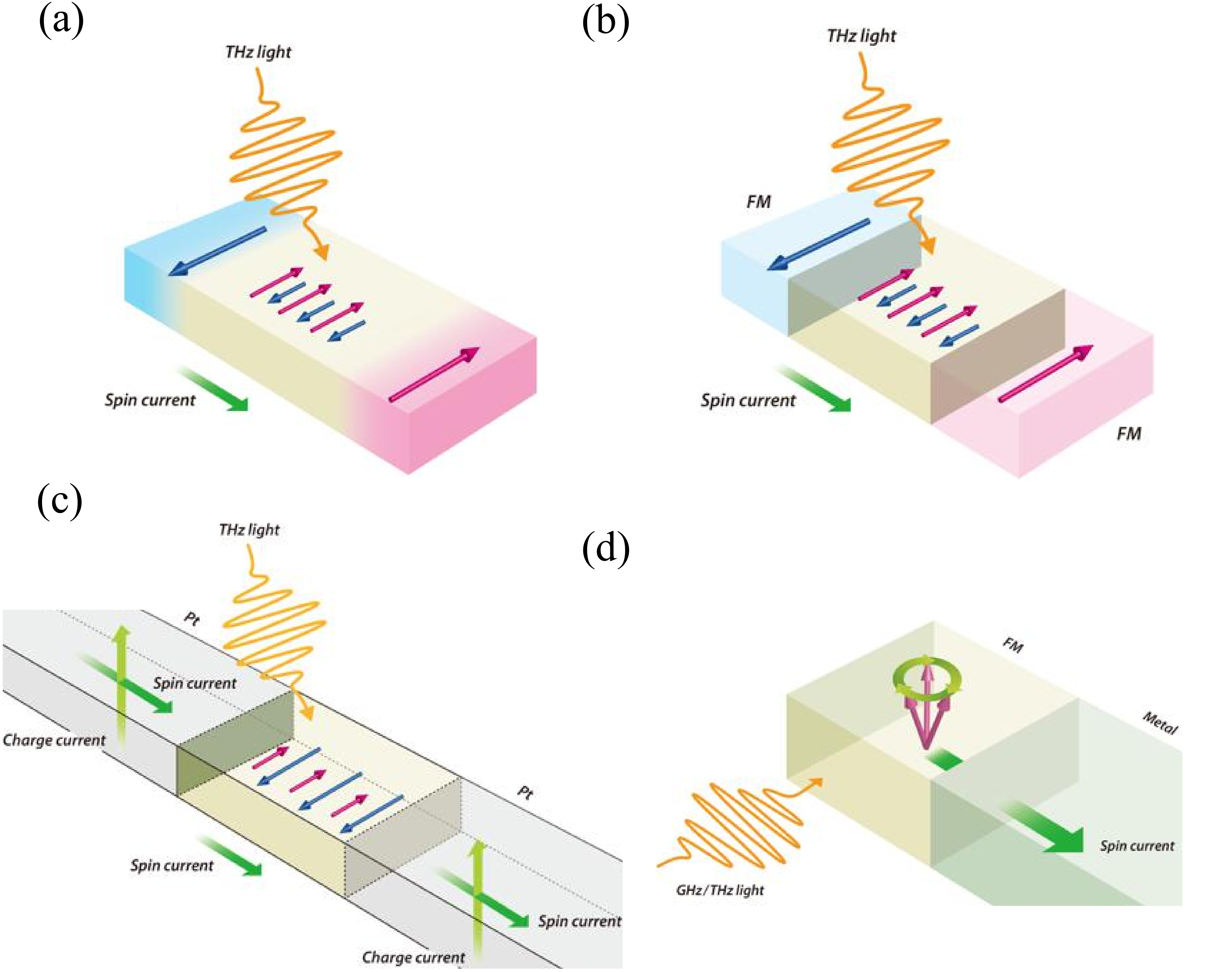}
  \caption{(Color online) Schematic figure of the experimental setups for measuring photo-induced spin current: All-optical setup [(a) and (b)] and two-terminal setup (c). (a) The all-optical setup irradiates the isolated magnet using THz light. The optically-induced spin current accumulates the angular momentum at the end of the magnet which is depicted by the clouds; it produces the asymmetric distribution of the angular momentum in the magnet. (b) A similar observation by attaching a thin layer of a soft ferromagnet at the two ends. The photovoltaic spin current is injected to or absorbed from the soft ferromagnets. (c) The two-terminal setup observes the directional flow of spin current using the inverse spin Hall effect. The optically-induced spin current flows along a certain direction of the system. Therefore, inverse spin Hall voltage of the two leads has the same sign. These setups are different from that of spin pumping of panel (d), in which a transverse AC field is applied to the magnet and the spin current is diffusively expanded.}
  \label{fig:experiment}
\end{figure}

%%%%%%%%%%%%%%%%%%%%%%%%%%%%%%%%%%%%%%%%%%

In this section, we discuss experimental methods for detecting signatures of a directional spin current in our mechanism.

%%%%%%%%%%%%%%%%%%%%%%%%%%%%%%%%%%%%%%%%%%%%
%%%%%%%%%%%%%%%%%%%%%%%%%%%%%%%%%%%%%%%%%%%%
%%%%%%%%%%%%%%%%%%%%%%%%%%%%%%%%%%%%%%%%%%%%
\subsection{Setup}

We here discuss experimental setups for the observation of the spin current generated by linearly-polarized light. The mechanism studied here produces a directional flow of the spin current, which is a distinct feature from the spin pumping~\cite{Kajiwara2010,Heinrich2011}. Therefore, the observation of the directional flow should provide an evidence for our mechanism. We discuss two different mechanisms: First one is an all-optical setup using Kerr rotation or Faraday effect, and the second is a two-terminal setup using inverse spin-Hall effect.

Observation of the spatial distribution of angular momentum in the open circuit setup provides a direct evidence for the optically-generated spin current [See Fig.~\ref{fig:experiment}(a)]. In an isolated magnet, the spin current produced by a THz light flows along a direction defined by the magnetic order and the crystal symmetry. Therefore, if the system becomes close enough to a laser-driven non-equilibrium steady state, the angular momentum accumulates at the two ends in an open circuit setup in Fig.~\ref{fig:experiment}(a); positive angular momentum on one end and negative on the other end. The angular momentum distribution is anti-symmetric along the direction of the spin current. This distribution is strikingly different from the spin pumping case in which the distribution is symmetric and its difference from the equilibrium state is larger at the focal area of the laser than at the ends. 

An all-optical setup using Kerr rotation or Faraday effect would be a useful setup for the observation of such a spatial distribution. Measurement of magnetic moments and its spatial distribution using the optical probe is a commonly used technique for observing the spin current. For instance, this method is used to observe the spin Hall effect~\cite{Kato2004}. Similarly, observing the magnetization of soft magnet layers attached to the two ends is another possible setup for the experiment [Fig.~\ref{fig:experiment}(b)]. 

The observation of spin current in a two-terminal setup in Fig.~\ref{fig:experiment}(c) also enables us to see the directional flow of spin current and to distinguish it from the spin pumping effect. This setup consists of a noncentrosymmetric magnetic insulator which is sandwiched between two metallic leads; the two leads detect spin current via inverse spin Hall effect~\cite{Saitoh2006,Valenzuela2006,Kimura2007}. In the photovoltaic mechanism, the spin current in the two leads flows toward the same direction. Therefore, the inverse spin Hall voltage of the two leads has the same sign. In contrast, in the spin pumping, the spin current diffusively flows outward from the magnet; the inverse spin Hall voltage is positive on one side and negative on the other. Therefore, the relative sign of the inverse spin Hall voltage of the two leads can make a distinction between the spin pump and our mechanism.

Finally, we shortly comments on heating effect of applied electromagnetic waves. When we try to detect the photovoltaic spin current with the above setups, spin pumping might also occur due to the heating effect of the applied laser. For such a case, extracting the asymmetric part of the angular-momentum distribution or inverse spin Hall voltage is important to detect an evidence for our mechanism.

%%%%%%%%%%%%%%%%%%%%%%%%%%%%%%%%%%%%%%%%%%%%
%%%%%%%%%%%%%%%%%%%%%%%%%%%%%%%%%%%%%%%%%%%%
%%%%%%%%%%%%%%%%%%%%%%%%%%%%%%%%%%%%%%%%%%%%
\subsection{Required intensity of AC field}

We next estimate the required ac electromagnetic field for generating an observable spin current. We here assume a spin current of $J_s=10^{-16}$ J/cm$^2$ is observable. This estimate is based on a Boltzmann theory calculation for spin Seebeck effect in a ferromagnet~\cite{Hirobe2017,Ishizuka2019}. The details of the estimate is briefly explained in Appendix~\ref{sec:Boltzmann}. We use the following parameters as a typical value for 1D insulating magnets: $J=100k_B$ J, $\delta=0.1$, $S_A=S_B=1$, $g_A-g_B=0.1\mu_B$ J/T, $h_+=10k_B$ J, and a the light with a frequency which is $\hbar\delta\omega=6\pi\hbar\times10^{11}$ Hz above the band gap. Here, $\hbar$ is the Planck constant. With these parameters, the conductivity for the 1D AFM/FRM chain is $\Re\left[\sigma(0;\omega,-\omega)\right]\sim 10^{-14}$ J/(cm$^2$T$^2$). Therefore, the required magnitude of oscillating magnetic field to produce a spin current of $J_s=10^{-16}$ J/cm$^2$ is $B\sim\sqrt{\frac{J_s}{|\Re\left[\sigma(0;\omega,-\omega)\right]|}}\sim 0.1$ T. This corresponds to the electric field $E=cB\sim 10^4-10^5$ V/cm under the assumption of $c=10^8$ m/s which is a typical value of speed of light in insulators. Similar estimate for the 3D magnet with $J=100k_B$ J, $J_\perp=10k_B$ J, $\delta=0.1$, $S_A=S_B=1$, $g_A-g_B=0.1\mu_B$ J/T, $h_+=10k_B$ J, and $\omega=2\pi\times10^{12}$ Hz gives $\Re\left[\sigma(0;\omega,-\omega)\right]\sim 10^{-11}$ J/(cm$^2$T$^2$) and $E=c B\sim10^5-10^6$ V/cm. Our estimate predicts that the photovoltaic spin current is experimentally observable by using a moderate-intensity THz light.

%%%%%%%%%%%%%%%%%%%%%%%%%%%%%%%%%%%%%%%%%%%%
%%%%%%%%%%%%%%%%%%%%%%%%%%%%%%%%%%%%%%%%%%%%
%%%%%%%%%%%%%%%%%%%%%%%%%%%%%%%%%%%%%%%%%%%%
\subsection{Candidate material}

We believe the photovoltaic spin current should be seen generically in noncentrosymmetric magnets. In a recent work~\cite{Ishizuka2019}, the authors find three kinds of spin-light couplings induce the spin current in a spin chain, and this work presents photovoltaic spin current in ordered magnets. These results imply the generation of photovoltaic spin current is a universal phenomenon in noncentrosymmetric magnetic insulators. One such material is ferrimagnetic diamond chains~\cite{Okamoto2003,Shores2005,Vilminot2006}. These materials often have a distortion associated with trimerization, which breaks the inversion symmetry~\cite{Okamoto2003}. Also, a large density of states for the magnon excitations is expected in this material because it is a quasi-1D magnet. Thus the ferrimagnetic phase of the diamond chain is a promising candidate for studying the spin current.

%%%%%%%%%%%%%%%%%%%%%%%%%%%%%%%%%%%%%%%%%%%%
%%%%%%%%%%%%%%%%%%%%%%%%%%%%%%%%%%%%%%%%%%%%
%%%%%%%%%%%%%%%%%%%%%%%%%%%%%%%%%%%%%%%%%%%%
\section{Summary and Discussion}\label{sec:summary}

To summarize, we studied the spin current generation through the shift current mechanism in ferrimagnetic/antiferromagnetic insulators. Our theory uses a nonlinear response theory, which is a natural generalization of the linear response theory. Based on this method, we find that the illumination of a linearly-polarized light produces the magnon current in noncentrosymmetric magnets with antiferromagnetic/ferrimagnetic order. The photovoltaic spin current appears even at the zero temperature where no magnon excitation exists in the equilibrium; the current is related to the two-magnon excitation process and not to the optical transition of existing (thermally-excited) magnons. We stress that the photo-induced spin current in our mechanism is carried by electrically-neutral particles. The relaxation-time dependence of the spin current indicates that our photovoltaic effect is a ``shift current'', i.e., the nonlinear conductivity is insensitive to the damping. Our theory clearly shows that the shift current mechanism, which is well known in electron (fermion) systems, is also relevant to systems with bosonic excitations whose the ground state is the vacuum of bosons (zero boson state).

Our result implies the zero-point quantum fluctuation is a key for the shift-current type photocurrent. In the spinon spin current~\cite{Ishizuka2019}, the optical transition of a fermionic excitation plays a crucial role for the photocurrent. In contrast to these cases, the ground state of the ordered magnets is the zero-magnon state. Therefore, no optical transition of the existing magnons. Despite the crucial difference, we find a finite photovoltaic spin current at the zero temperature. The magnon photocurrent we found is ascribed to the optical transition of the ``condensed'' Holstein-Primakov bosons. In the antiferromagnets/ferrimagnets, the ground state is a condensate of Holstein-Primakov bosons, which is technically represented by the Bogoliubov transformation. The optical transition of the condensed Holstein-Primakov bosons allows generation of the shift-current type photocurrent even at the zero temperature. On the other hand, we find that the nonlinear conductivity is zero at $T=0$ for the ferromagnetic version of the model considered here. From this viewpoint, the two-magnon creation is similar to the particle-hole pair creation in semiconductors; the optical transition of fermions from the valence band to the conduction band is equivalent to the pair creation. As the condensation of the Holstein-Primakov bosons is a manifestation of zero-point fluctuation, the zero-point fluctuation is the essence for the shift-current type photovoltaic effects in the magnetic insulators. 

Our results implies that the dc spin current generation using linearly polarized light is generally possible in the magnets without inversion symmetry.

%%%%%%%%%%%%%%%%%%%%%%%%%%%%%%%%%%%%%%%%%%%%
%%%%%%%%%%%%%%%%%%%%%%%%%%%%%%%%%%%%%%%%%%%%
%%%%%%%%%%%%%%%%%%%%%%%%%%%%%%%%%%%%%%%%%%%%
\appendix
\section{Derivation of Kraut-von Baltz formula for Bosons}\label{sec:ResponseEq}

Here, we shortly explain the derivation of the nonlinear conductivity in two-band boson systems. We used the formula in Eq.~\eqref{eq:formula:pair} for the analytic calculations and Eq.~\eqref{eq:formula:BdG} for numerical results with a finite Gilbert damping.

We calculate the nonlinear response coefficients using a formalism similar to the linear response theory. We assume a system with a time-dependent perturbation $H'=-\sum_\mu \hat B_\mu F_\mu(t)$, where $\hat B_\mu$ is an operator and $F_\mu(t)$ is a time-dependent field; the Hamiltonian reads $H=H_0+H'$. The expectation value of an observable $\hat A$ reads $\langle\hat A\rangle(t) = {\rm Tr}\left[\hat \rho(t)\hat A\right]/Z,$ where $\rho(t)$ is the density matrix at time $t$ and $Z\equiv{\rm Tr}\rho(t)$. By expanding $\rho(t)$ up to the second order in $F_\mu(t)$, the Fourier transform of $\langle A\rangle(t)$, $\langle A\rangle(\Omega)$, reads
\begin{align}
\langle A\rangle(\Omega)=\sum_{\mu,\nu}\int d\omega\,\sigma_{\mu\nu}(\Omega;\omega,\Omega-\omega)F_{\mu}(\omega)F_{\nu}(\Omega-\omega),
\end{align}
with the nonlinear conductivity
\begin{align}
\sigma_{\mu\nu}&(\Omega;\omega,\Omega-\omega)=\frac1{2\pi}\sum_{n,m,l}\frac{(\rho_n-\rho_m)(B_\mu)_{nm}}{\omega-E_m+E_n-{\rm i}/(2\tau_{mn})}\nonumber\\
&\times\left[\frac{(B_\nu)_{ml}A_{ln}}{\Omega+E_n-E_l-{\rm i}/(2\tau_{mn})}-\frac{A_{ml}(B_\nu)_{ln}}{\Omega+E_l-E_m-{\rm i}/(2\tau_{mn})}\right].\label{eq:ResponseEq:sigma1}
\end{align}
Here, $E_n$ is the eigenenergy of the many-body eigenstate $n$, $\tau_{mn}$ is the relaxation time, and $O_{nm}$ ($O=A,B_\mu,B_\nu$) is the matrix element of $\hat O$ in the eigenstate basis of $H_0$.

We here consider a periodic free-boson system in which all matrices $A$, $B^\mu$, and $B^\nu$ have the following form:
\begin{align}
\hat O=&\sum_{\bm k}
\left(\begin{array}{cc}
\alpha^\dagger_{\bm k} & \beta_{-\bm k}
\end{array}\right)
O_{\bm k}
\left(\begin{array}{c}
\alpha_{\bm k} \\
\beta_{-\bm k}^\dagger
\end{array}\right),\\
=&\sum_{\bm k}
\left(\begin{array}{cc}
\alpha^\dagger_{\bm k} & \beta_{-\bm k}
\end{array}\right)
\left(\begin{array}{cc}
(O_{\bm k})_{\alpha\alpha} & (O_{\bm k})_{\alpha\beta} \\
(O_{\bm k})_{\beta\alpha} & (O_{\bm k})_{\beta\beta} 
\end{array}\right)
\left(\begin{array}{c}
\alpha_{\bm k} \\
\beta_{-\bm k}^\dagger
\end{array}\right),
\end{align}
where $\alpha_{\bm k}$ ($\alpha_{\bm k}^\dagger$) and $\beta_{\bm k}$ ($\beta_{\bm k}^\dagger$) are the annihilation (creation) operators of the boson eigenstates with momentum $\bm k$, and $O_{\bm k}=A_{\bm k}, B^\mu_{\bm k}, B^\nu_{\bm k}$. The theory for spinwave excitations of many antiferromagnetic models with a N\'eel-type order reduces to the above form by using Holstein-Primakov and Bogoliubov transformations.

For the two-band system, we can express Eq.~\eqref{eq:ResponseEq:sigma1} using single-particle eigenstates. We note that $A$, $B_\mu$, and $B_\nu$ for the two-band system above do not conserve the particle number. However, all operators are quadratic in the annihilation/creation operators and consists of only for terms: $\alpha_{\bm k}^\dagger\alpha_{\bm k}$, $\beta_{-\bm k}\beta_{-\bm k}^\dagger$, $\beta_{-\bm k}\alpha_{\bm k}$, and $\alpha_{\bm k}^\dagger\beta_{-\bm k}^\dagger$. Therefore, only few terms out of the possible Wick decomposition remain nonzero, similar to that of the systems with conserved particle number. Using these features, we find
\begin{align}
&\sigma(\Omega;\omega,\Omega-\omega)=\nonumber\\
&\frac1{2\pi} \sum_{\bm k,a_i=\alpha,\beta}\frac{{\rm sgn}(a_3)(\tilde\rho_{\bm k,a_1}{\rm sgn}(a_2)-{\rm sgn}(a_1)\tilde\rho_{\bm k,a_2})(B^\mu_{\bm k})_{a_1a_2}}{\omega-\tilde\varepsilon_{a_2}(\bm k)+\tilde\varepsilon_{a_1}(\bm k)-i/(2\tau_{\bm k})}\nonumber\\
&\qquad\qquad\times\left[\frac{(B^\nu_{\bm k})_{a_2a_3}(A_{\bm k})_{a_3a_1}}{\Omega+\tilde\varepsilon_{a_1}(\bm k)-\tilde\varepsilon_{a_3}(\bm k)-i/(2\tau_{\bm k})}\right.\nonumber\\
&\qquad\qquad\qquad\left.-\frac{(A_{\bm k})_{a_2a_3}(B^\nu_{\bm k})_{a_3a_1}}{\Omega+\tilde\varepsilon_{a_3}(\bm k)-\tilde\varepsilon_{a_2}(\bm k)-i/(2\tau_{\bm k})}\right].\label{eq:formula:BdG}
\end{align}
Here,
\begin{align}
{\rm sgn}(a)=&\left\{\begin{array}{rl}
1 & (a=\alpha)\\
-1& (a=\beta)
\end{array}\right.,\\
\tilde\varepsilon_{a}(\bm k)=&{\rm sgn}(a)\varepsilon_{a}(\bm k),\\
\tilde\rho_{\bm k,a}=&\left\{\begin{array}{rl}
\langle\alpha^\dagger_{\bm k}\alpha_{\bm k}\rangle_0 & (a=\alpha)\\
\langle\beta_{-\bm k}\beta_{-\bm k}^\dagger\rangle_0& (a=\beta)
\end{array}\right.,
\end{align}
and we assumed the relaxation time only depends on $\bm k$. It is worth noting that the conductivity remains finite at $T=0$ despite there are no excitations. Technically, this is a consequence of $\tilde\rho_{\bm k,\beta}$, which is 1 at $T=0$. Physically, this is because the pair creation/annihilation processes contribute to the spin current even at $T=0$. 

We here focus on the $T=0$ limit. In this limit, $\tilde\rho_{\bm k,\alpha}=0$ and $\tilde\rho_{\bm k,\beta}=1$. Using these results, we obtain
\begin{widetext}
\begin{align}
&\sigma(0;\omega,-\omega)=-\frac1\pi\sum_{\bm k,a_i=\alpha,\beta}\left[\frac{(1+i2\tau\omega)|B_{\beta\alpha}|^2(A_{\alpha\alpha}+A_{\beta\beta})}{(\omega-i/2\tau_{\bm k})^2-(\varepsilon_{\alpha}(\bm k)+\varepsilon_{\beta}(\bm k))^2}\right]\nonumber\\
&+\frac1{2\pi}\sum_{\bm k,a_i=\alpha,\beta}\left\{\frac{(B_{\bm k})_{\beta\alpha}(A_{\bm k})_{\alpha\beta}((B_{\bm k})_{\beta\beta}+(B_{\bm k})_{\alpha\alpha})}{(\omega-i/2\tau_{\bm k}-\varepsilon_\alpha(\bm k)-\varepsilon_\beta(\bm k))(\varepsilon_\alpha(\bm k)+\varepsilon_\beta(\bm k)+i/2\tau_{\bm k})}\right.\nonumber\\
&\left.+\frac{(B_{\bm k})_{\alpha\beta}(A_{\bm k})_{\beta\alpha}((B_{\bm k})_{\beta\beta}+(B_{\bm k})_{\alpha\alpha})}{(\omega-i/2\tau_{\bm k}+\varepsilon_\alpha(\bm k)+\varepsilon_\beta(\bm k))(\varepsilon_\alpha(\bm k)+\varepsilon_\beta(\bm k)-i/2\tau_{\bm k})}\right\}.
\end{align}
\end{widetext}
As we discussed in the main text, certain symmetries restricts the first term to be zero; this is the case for the models we consider in the main text. Assuming the first term vanishes, we find
\begin{align}
\Re&\left[\sigma(0;\omega,-\omega)\right]=\nonumber\\
&-\frac1{\pi}\Re\left\{\frac{(B_{\bm k})_{\beta\alpha}(A_{\bm k})_{\alpha\beta}((B_{\bm k})_{\beta\beta}+(B_{\bm k})_{\alpha\alpha})}{\omega^2-(\varepsilon_\alpha(\bm k)+\varepsilon_\beta(\bm k)+i/2\tau)^2}\right\}.\label{eq:formula:pair}
\end{align}
We used this formula for the calculation of nonlinear conductivity in the main text.

%%%%%%%%%%%%%%%%%%%%%%%%%%%%%%%%%%%%%%%%%%%%
%%%%%%%%%%%%%%%%%%%%%%%%%%%%%%%%%%%%%%%%%%%%
%%%%%%%%%%%%%%%%%%%%%%%%%%%%%%%%%%%%%%%%%%%%
\section{Boltzmann theory for spin Seebeck effect}\label{sec:Boltzmann}

The magnitude of spin current $J_s=10^{-16}$ J/cm$^2$ is the estimate for the spinon spin current produced by the spin Seebeck effect in a recent experiment~\cite{Hirobe2017}. We here summarize the method and result discussed in a supplemental material of a recent work~\cite{Ishizuka2019}.

The spin current is estimated from the Seebeck effect of magnons whose dispersion is given by
\begin{align}
\varepsilon(\bm k)=JSk^2+2DS+h.
\end{align}
near the $\Gamma$ point of $\bm k=\bm 0$. This magnon dispersion corresponds to that of a ferromagnetic heisenberg model with exchange interaction $J$, uniaxial anisotropy $D$, and the magnetic field $h$ parallel to the anisotropy. The current is calculated using the semiclassical Boltzmann theory, in which the current reads
\begin{align}
J_{s}(\bm r)=\hbar\int \frac{d\bm k}{(2\pi)^3} v_z f_{\bm k}(\bm r). \label{eq:Boltzmann:Js}
\end{align}
Here, $f_{\bm k}(\bm r)$ is the density of magnons with momentum $\bm k$ at position $\bm r$ and $v_z\equiv \partial_{k_z}\varepsilon(\bm k)$ is the group velocity of magnons. $f_{\bm k}(\bm r)$ is calculated from the Boltzmann equation with temperature gradient
\begin{align}
\bm v_{\bm k}\cdot\bm \nabla_r f_{\bm k}(\bm r)=-\frac{f_{\bm k}(\bm r)-f^{(0)}_{\bm k}(\bm r)}{\tau_{\bm k}},\label{eq:Boltzmann:Boltzmann}
\end{align}
where $f^{(0)}_{\bm k}(\bm r)$ is the density at the equilibrium. Here, the relaxation-time approximation is used to simplify the calculation of collision integral on the right hand side. The spin current induced by the spin Seebeck effect is estimated by substituting the solution of $f_{\bm k}(\bm r)$ in Eq.~\eqref{eq:Boltzmann:Boltzmann} into the current formula in Eq.~\eqref{eq:Boltzmann:Js}

In the Boltzmann theory, the spin current by the spin Seebeck effect reads
\begin{align}
J_{s}(\bm r)\sim&\frac{3(6\pi^2)^{\frac23}J_H^2S^2}{2\alpha k_B a T(\bm r)}
\frac{\Delta T}{T(\bm r)}
F\left(\frac{J_HSa^2\Lambda^2}{2k_B T(\bm r)},\frac{2DS+h}{2k_B T(\bm r)}\right),
\end{align}
where $\Lambda=(6\pi^2)^{1/3}/a$ is the cutoff for magnon dispersion and
\begin{align}
F(a,b)= \int_0^1x^4{\rm csch}^2(ax^2+b)dx.
\end{align}
Using a set of typical parameters $S=1$, $J_H=100k_B$ J, $D=0$ J, $h=\mu_B$ J, $a=4\times10^{-10}$ m, $\alpha=10^{-2}$, $T=100$ K, $\Delta T=3\times10^4$ K/m, we find $J_{s}\sim 10^{-12}$ J/cm$^2$ for the ferromagnet. We assume this value as the typical spin current density in the insulating ferromagnets.

A recent experiment on quasi-one-dimensional magnets observed a spin current which is 10$^{-4}$ of what is typically observed in a ferromagnetic phase~\cite{Hirobe2017}. Therefore, we assume $J_{s}\sim 10^{-16}$ J/cm$^2$ as the experimental resolution for the spin current.

\acknowledgements
We thank Ryosuke Matsunaga and Youtarou Takahashi for fruitful discussions. We also thank Wataru Murata for providing Fig.~\ref{fig:experiment}. H.I. was supported by JSPS KAKENHI Grant Numbers JP18H04222, JP19K14649, and JP18H03676, and CREST JST Grant Numbers JPMJCR16F1. M.S. was supported by JSPS KAKENHI (JP17K05513), and Grant-in-Aid for Scientific Research on Innovative Area ``Nano Spin Conversion Science'' (Grant No.17H05174) and ``Physical Properties of Quantum Liquid Crystals'' (Grant No. 19H05825).

%%%%%%%%%%%%%%%%%%%%%%%%%%%%%%%%%%%%%%%%%%%%%%%%%%%%

\end{document}